\documentclass{article}
\begin{document}
\title{\large \bf BRANE-UNIVERSE IN SIX DIMENSIONS}
\author{
{\bf Merab Gogberashvili$^a$ and Paul Midodashvili$^b$}\\
$^a$ Andronikashvili Institute of Physics, 6 Tamarashvili Str.,
Tbilisi 380077, Georgia \\
{\sl E-mail: gogber@hotmail.com }\\
$^b$ Tskhinvali State University, 1 Cherkezishvili Str.,
Gori 383500, Georgia \\
{\sl E-mail: midodashvili@hotmail.com }
}
\maketitle
\begin{center}
\quotation{
Growing non-singular solution of 6-dimensional Einstein equations for the
4-brane in infinite transversal 2-space is found. This solution provides
gravitational trapping of matter and 4-dimensional gravity on the brane
without extra delta-like source. The suitable solution exists in the case
of the (2+4)-space and not exists for the (1+5)-signature.
}
\end{center}
\vskip 1.0cm

%%%%%%%%%%%%%%%%%%%%%%%%%%%%%%%%%%%%%
After the papers \cite{ADD,RaSu,G1} the models with brane-Universe in a
high dimensional space-time have become popular. Large extra dimensions
may play an important role in solving such problems as smallness of
cosmological constant, the origin of the hierarchy, the nature of flavor,
the source of SUSY breaking, etc. As in previous articles
\cite{G1,G2,G3,G4,G5}, in the present paper the Universe is considered as
a single brane.

Brane models need some natural matter localization mechanism and
explanation of 4-dimensional Newton's law. The question of matter
localization on the brane has been investigated in various papers (for
recent ideas see \cite{DvSh}). In our opinion the localizing force must
be universal for all types of 4-dimensional matter fields. In our world
gravity is known to be the unique interaction which has universal coupling
with all matter fields. If extended extra dimensions exist, it is natural
to assume that trapping of matter on the brane has a gravitational nature.
Because of problems with quantum field theory in curve space it is
difficult to find exact trapping mechanism for different matter fields
separately. Our approach is more general: to provide universal and stable
trapping we assume that on the brane (where all gravitating matter can
reside) gravitational potential should have minimal value with respect to
extra coordinates.

Growing gravitational potential is the opposite choice compared to the
one of Randall-Sundrum with the maximum on the brane \cite{RaSu}. We
assume that 4-dimensional Newton's law on the brane is the result of the
cancellation mechanism introduced in \cite{G1,G4} (as it is in
Randall-Sundrum's case also), which allows both types of gravitational
potential. This mechanism can be used even for a simple 5-dimensional model
with exponentially growing transversal gravitational potential. Far from
the brane (natural cutoff is the width of the brane) transversal
gravitational potential becomes large and usual splitting of 5-dimensional
space of \cite{RaSu,G1,G2,G3} cannot be valid any more. This is similar to
the case of QCD where nobody worries about the increasing quark potential.

In this paper we introduce a realistic model of gravity and matter
localization on (1+3)-brane embedded in a (2+4)-space in the case of
growing transversal gravitational potential. Our motivation for the
choice of the signature is as follows. In the massless field case (weakest
coupling with gravity) symmetries of a multi-dimensional manifold can be
restored. It is well known, that in the zero-mass limit the main equations
of physics are invariant under the 15-parameter nonlinear conformal
transformations. A long time ago it was also discovered that the conformal
group can be written as a linear Lorentz-type transformation in a
(2+4)-space (for these subjects see for example \cite{PeRi}).

The (2+4)-space was earlier investigated in \cite{G5}, but for a (2+3)-brane
with two open time directions. The case of (1+3)-brane in six dimensions had
been considered by different authors \cite{Su,ChPo,CoKa,Gr,GhSh}. In all
these papers space-like extra dimensions were investigated. We will show 
that
a suitable solution of our minimal model with growing gravitational 
potential
does not exist for the case of (1+5)-space.

In all (1+5)-models the authors used polar coordinates for the transversal
2-space. At the origin, where the brane is placed, polar coordinates are
singular and the choice of acceptable boundary conditions is problematic.
In our approach we consider transversal (1+1)-space and use nonsingular
Cartesian coordinates.

Einstein's equations in six dimension with a bulk cosmological constant
$\Lambda$ and stress-energy tensor $T_{AB}$ can be written in the form
\begin{equation} \label{1}
^6R_{AB} = -\frac {1}{2}\Lambda g_{AB} + \frac {1}{M^4}\left( T_{AB}-
\frac {1}{4} g_{AB}T\right),
\end{equation}
where $^6R_{AB}$ and $M$ are respectively the Ricci tensor and the
fundamental scale. Capital Latin indices run over $A, B,... = 0, 1, 2, 3,
5, 6 $.

We are looking for a solution of (\ref{1}) in the form
\begin{equation} \label{2}
ds^2= \phi ^2(z) \eta_{\alpha \beta }(x^\nu)dx^\alpha dx^\beta +
g_{ij}(z)dx^i dx^j,
\end{equation}
providing stable splitting of the sub-manifold \cite{G3}. Greek indices
$\alpha, \beta,... = 0, 1, 2, 3$ numerate coordinates in 4-dimensions,
while small Latin indices $i, j, ... = 5, 6$ - coordinates of the
transversal space.

In (\ref{2}) only the 4-dimensional conformal factor $\phi ^2$ and the
metric tensor of transversal (1+1)-space $g_{ij}$ depends on the extra
coordinates $x^i$ via the dimensionless coordinate
\begin{equation} \label{3}
z = \frac{x^2_5 - x^2_6}{\epsilon^2} \ge 0,
\end{equation}
where $\epsilon $ is the width of the brane.

Suppose that the system of Einstein and matter field equations have the
solution with the minimal energy when extra coordinates enter stress-energy
from the metric (\ref{2}) only \cite{G3}. This means that strength of gauge
fields towards the extra directions and covariant derivations of matter
fields with respect to extra coordinates are zero $F^a_{iA} = D_i\psi^a = 0$
\cite{G3}. Then multi-dimensional matter energy-momentum tensor can be
written in the form
\begin{equation} \label{4}
T_{\alpha\beta } = \frac{\tau_{\alpha\beta }(x^\nu)}{\rho ^2
\phi ^2(z)}, ~~~~ T_{ij} = - g_{ij}(z)\frac{L(x^\nu)}{\rho^2 \phi ^4(z)} .
\end{equation}
The Lagrangian of matter fields $L(x^\nu)$ and the 4-dimensional
stress-energy $\tau_{\alpha \beta}(x^\nu )$ in the space (\ref{2})
automatically appears to be independent of $z$ (see for example 
\cite{PeRi}).

In (\ref{4}) $\rho $ is an arbitrary length scale. The only length scale of
our task is the width of the brane and it is natural to take
\begin{equation} \label{5}
\epsilon^2 = \rho^2.
\end{equation}

To localize the matter on the brane without extra sources the factor
$1/\phi ^2(z)$ in (\ref{4}) should have a delta-like behavior. It means
that $\phi^2(z)$ (and transversal gravitational potential) must be a
growing function starting from the brane location. Let us place our world
at the origin $z=0$, corresponding to a (1+3)-brane moving with the speed
of light in the transversal (1+1)-space.

Using the metric ansatz (\ref{2}) one can derive the following
decomposition of 6-dimensional Einstein's equations
\begin{eqnarray} \label{6}
R_{\alpha\beta} - \frac{1}{2}\eta_{\alpha\beta}\left( D_iD^i\phi ^2+
\frac{1}{2\phi ^2}D_i\phi ^2D^i\phi ^2\right)
= - \frac{\Lambda }{2}\phi^2 \eta_{\alpha\beta}
+ \frac{1}{\epsilon^2 M^4\phi^2 }\left( \tau
_{\alpha\beta} - \frac{\tau - 2L}{4}\eta_{\alpha\beta}\right), \nonumber \\
R_{ij} - \frac{2}{\phi ^2}\left( D_iD_j\phi ^2- \frac{1}{2\phi
^2}D_i\phi ^2 D_j\phi ^2\right)
= g_{ij}\left( - \frac{\Lambda }{2} - \frac{\tau + 2L}{4\epsilon^2
M^4\phi^4}\right),
\end{eqnarray}
where $R_{ij}$ and $D_i$ are respectively the Ricci tensor and the
covariant derivative containing the metric tensor $g_{ij}$ of the 2-space.
The Ricci tensor in four dimensions $R_{\alpha\beta}$ is constructed by
the 4-dimensional metric tensor $\eta_{\alpha\beta}(x^{\nu})$ in the
standard way.

Since the gravity in the 2-space is trivial, we have only one
independent component of $g_{ij}$. We choose
\begin{equation} \label{7}
g_{ij} = \eta _{ij}g(z),
\end{equation}
where $\eta _{ij}$ is the metric tensor of the flat (1+1)-space.

On the brane we require to have 4-dimensional Einstein equations without
a cosmological term
\begin{equation} \label{8}
R_{\alpha\beta} = \frac{1}{\epsilon^2 M^4\phi^2 }\left( \tau
_{\alpha\beta} - \frac{1}{2}\eta_{\alpha\beta}\tau\right).
\end{equation}
Then (\ref{6}) takes the form
\begin{eqnarray} \label{9}
z(\phi \phi ^{''}+3\phi ^{'2}) + \phi \phi ^{'} = \left(\phi^2 \Lambda -
\frac{\tau + 2L}{2\epsilon^2 M^4\phi^2 } \right)
\frac{\epsilon^2}{8}g, \nonumber \\
\frac{\phi^{''}}{\phi} = \frac{g^{'}}{g}, ~~~~~~~~~~~~~~~~~~~~~~\\
z\left( \frac{g^{''}}{g} - \frac{g^{'2}}{g^2} +
\frac{4\phi^{'}}{\phi}\frac{g^{'}}{g}\right)
+\frac{g^{'}}{g} + \frac{4\phi^{'}}{\phi}
=\left(\phi^2 \Lambda
+ \frac{\tau + 2L}{2\epsilon^2 M^4\phi^2 } \right) \frac{\epsilon^2
}{4\phi^2}g,
\nonumber
\end{eqnarray}
where a prime denotes a derivative with respect to $z$. We note that the
system (\ref{9}) is self-consistent only for strictly correlated extra
coordinates in the form of (\ref{3}), corresponding to the light-like
(1+3)-brane in (1+1)-space.

The solution of the second equation of the system (\ref{9}) ($5$ - $6$
component of Einstein equations) is
\begin{equation} \label{10}
g = c \phi^{'},
\end{equation}
where $c$ is an integration constant.

Because of our choice of the source (\ref{4}) the last equation of
(\ref{9}) is satisfied identically and only the first equation of the
system remains to be solved. The first integral of the remaining equation
can be written in the form
\begin{equation} \label{11}
z\phi^3 \phi^{'} + A\phi^5 + B \phi +C = 0,
\end{equation}
where $C$ is the second integration constant and the dimensionless
parameters
\begin{equation} \label{12}
A = - \frac{\Lambda \epsilon^2 c}{40},~~~~~ B = \frac{c(\tau + 2L)}{16 M^4} 
,
\end{equation}
were introduced. In general $B$ depend on the 4-coordinates $x^\nu$.

Using (\ref{2}) and (\ref{10}), integration of (\ref{8}) by the extra
coordinates gives the relation
\begin{equation} \label{13}
m^2_P = \frac{\int \sqrt {-g}dx_5dx_6}{\int \frac{1}{\epsilon^2
M^4\phi^2}\sqrt {-g}dx_5dx_6} =
\epsilon^2 M^4\frac{\int \phi^4d\phi}{\int\phi^2d\phi}
\end{equation}
between Planck's scale $m_P$ and the fundamental scale $M$. To have the
finite-volume integrals in (\ref{13}) in the infinite transversal 2-space
we need the solution of (\ref{11}) which approaches some finite value at
infinity. For simplicity, in this paper we consider only positive $\phi $,
since it enters the metric tensor quadratically. In order to avoid
singularity the function $\phi$ and its first derivative $\phi' > 0$,
which, according to (\ref{10}) enters in the metric of transversal space,
must be finite and nonzero on the brane.

In the case of pure gravitational field, when $B = 0$, there is no
continuous solution of (\ref{11}), which runs from the origin to
$z \rightarrow \infty $. At the point $z=1$ (corresponding, according to
(\ref{3}), to the brane-width equal to $\epsilon $ ) $\phi(z)$ becomes
infinitely large and in fact the transversal 2-space is closed. The brane
has the horizon of the size $\epsilon$ towards the extra dimensions,
similar to the interior of a black hole. So, 4-dimensional gravity with
ordinary Einstein equations (\ref{8}) is trapped on the brane itself by
multi-dimensional gravitation. This is a nonlinear effect of the
gravitational field.

When matter sources are added situation changes and there exists a 
nonsingular
solution of (\ref{11}) growing from the origin to some finite value at the
infinity. It means that 4-dimensional matter, which enters (\ref{11}) via 
the
parameter $B$, besides acting on the 4-dimensional curvature, also changes 
the
shape of the gravitational potential in the transversal space.

Boundary conditions are taken in the form
\begin{equation} \label{14}
\phi (z \rightarrow 0) \approx 1 + \frac{z}{|c|},~~~~~\phi (z \rightarrow
\infty)\approx a - \frac{1}{b |c| z^b}
\end{equation}
where $a > 1$ is the value of $\phi$ at the infinity and $b > 0$. This
choice corresponds to the following geometry of the bulk space-time on
the brane and in the transversal infinity:
\begin{eqnarray} \label{15}
ds^2(z \rightarrow 0) \approx \eta_{\alpha \beta }(x^\nu)dx^\alpha
dx^\beta + \eta_{ij}dx^i dx^j, \nonumber \\
ds^2 (z \rightarrow \infty)
\approx a^2 \eta_{\alpha \beta }(x^\nu)dx^\alpha dx^\beta +
\frac{1}{z^{b + 1}}\eta_{ij}dx^i dx^j.
\end{eqnarray}
At the origin $\phi $ is assumed to be equal to $1$, since any other
integration constant in (\ref{14}) will correspond to an overall rescaling
of the coordinates $x^A$ in (\ref{15}).

Conditions (\ref{14}) impose certain relations
\begin{eqnarray} \label{16}
Aa^5 + Ba + C \approx 0,~~~~~A + B + C \approx 0, \nonumber \\
b a^3 - 5Aa^4 - B \approx 0,~~~~~1 + 5A + B \approx 0.
\end{eqnarray}
>From these relations one can find
\begin{eqnarray} \label{17}
b \approx \frac{4a^3 + 3a^2 + 2a +1}{a^3(a^3 + 2a^2 + 3a + 4)},
\nonumber \\
A = -\frac{\Lambda \epsilon^2c}{40}\approx \frac{1}{a^4 + a^3 + a^2 + a
- 4}, \\
B = \frac{c(\tau + 2L)}{16 M^4} \approx - \frac{a^4+a^3 + a^2 + a +1}{a^4 + 
a^3
+ a^2 + a - 4}. \nonumber
\end{eqnarray}

For the realistic (similar to \cite{ADD}) values of our physical parameters
\begin{equation} \label{18}
m^2_P >> M^4\epsilon^2, ~~~~~(\tau + 2L) \sim M^4 > 0,
\end{equation}
from the relations (\ref{13}) and (\ref{17}) follows
\begin{eqnarray} \label{19}
a >> 1,~~~~~~~~~c \sim - 10,~~~~~~~~~\Lambda > 0,~ \nonumber \\
b \sim \frac{1}{a^3},~~~~~\epsilon^2 \sim \frac{1}{\Lambda a^4},~~~~~m^2_P
\sim M^4\epsilon^2 a^2.
\end{eqnarray}
So, smallness of the 4-dimensional gravitational constant $\sim 1/m^2_P$
and of the width of our world $\sim \epsilon $ can be the result of a
large values of the transversal gravitational potential $a$ and bulk
cosmological constant $\Lambda $.

Since $c$ is negative and $\phi'$ is positive, as it is seen from 
(\ref{10}),
a suitable solution of our model does not exist in the case of
space-like transversal 2-space studied in \cite{Su,ChPo,CoKa,Gr,GhSh}. In
the present paper we do not consider a pure time-like 2-space.

Using the relations (\ref{16}) one can show that the solution of
(\ref{11}) has an inflection point on the brane $z = 0$ (at the inflection
point second derivative of a function is zero, while the first is not). It
means that the transversal curvature $R_{zz}$ is zero on the brane, at the
minimum of the transversal gravitational potential. The function $\phi $
has no other inflection point outside the brane and smoothly grows from
$1$ to its maximal value $a$.

To summarize, in this paper it is shown that for the realistic values of
the fundamental scale and brane stress-energy, there exists a non-singular
static solution of (2+4)-dimensional Einstein equations corresponding to
infinite transversal (1+1)-space. This solution provides gravitational
trapping of the 4-dimensional gravity and the matter on the brane
without extra sources. In contrast to Randall-Sundrum's case, the factor
responsible for this trapping is the growing function away from the brane,
but has a convergent volume integral, although the transversal 2-space is
infinite. In our model any point-like particle in four dimensions can be a
(1+1)-dimensional object and there is possibility for nontrivial
applications of the classical theory of the strings.

{\bf Acknowledgements:} M. G. would like to acknowledged the hospitality
extended during his visits at the High Energy Physics Divisions of
Helsinki University and of Abdus Salam International Centre for
Theoretical Physics where this work was completed.

This work partially supported by the Academy of Finland under the
Project No. 163394.

\end{document}